\documentclass[aps,pre,twocolumn,superscriptaddress,floats,floatfix]{revtex4-2}
\usepackage{amssymb,amsmath,amsthm}
\usepackage{graphicx}
\usepackage{epstopdf}
\usepackage{color}
\usepackage{subfigure}
\usepackage{epsfig}
\usepackage{rotating}
\usepackage{mwe}
\usepackage{float}
\usepackage{dcolumn,bm}
\usepackage{verbatim}
\usepackage{hyperref}
\usepackage{pgf,tikz}
\usetikzlibrary{calc}
\usetikzlibrary{arrows.meta, angles, quotes}
\usepackage{makecell}
\usetikzlibrary{3d}
\usepackage[normalem]{ulem}
\usepackage{xcolor}
\usepackage{soul}
\usepackage{lineno}
\usepackage[dvipsnames]{xcolor}

\definecolor{Darkred}{RGB}{180, 0, 0}
\definecolor{Darkblue}{RGB}{26,68, 171}
\hypersetup{
  colorlinks = true,     
  urlcolor   = Darkblue, 
  linkcolor  = Darkblue, 
  citecolor  = Darkred   
}

\begin{document}

\begin{abstract}
The breakup of thinning (stretching) liquid filaments is strongly influenced by localized perturbations arising from impurities or suspended particles. Using numerical simulations and analytical modelling, we investigate the role of a solid particle on the breakup dynamics of a stretching liquid filament. We show that particle-induced perturbations trigger a universal pinch-off dynamics in the viscous regime. Once the filament surface approaches the particle, the subsequent breakup becomes self-similar and independent of the particle size. We derive an analytical expression for the pinch-off time based on the interplay between filament stretching and Rayleigh–Plateau instability, which agrees quantitatively with simulations. Our results reveal a universal mechanism by which localized perturbations control the breakup of viscous filaments containing solid particles.
\end{abstract}
\title{Self-similar breakup of a viscous liquid filament with a solid particle}

\author{Sanjay Shukla}
\email{s.shukla@tue.nl}
\affiliation{Department of Applied Physics and Science Education, Eindhoven University of Technology, 5600 MB Eindhoven, The Netherlands}
\author{Federico Toschi}
\email{F.Toschi@tue.nl}
\affiliation{Department of Applied Physics and Science Education, Eindhoven University of Technology, 5600 MB Eindhoven, The Netherlands}

\maketitle

\par{\textit{Introduction}-}
Thin liquid filaments and their stretching arise in many natural and industrial processes, including spray formation~\cite{Lin_1998_annurev.fluid}, inkjet printing~\cite{ARDEKANI_2010_JFM, Fraters_2020_PRA,Segers_2021_JFM}, and droplet generation in microfluidic devices~\cite{Todd_2001_PRL,laure_2006_PRE}. Their three-dimensional (3D) breakup is governed by the Rayleigh--Plateau instability~\cite{Plateau_1849,Rayleigh_1879,Rayleigh_1892}, in which capillary forces amplify the most unstable surface perturbation until the filament pinches off into droplets~\cite{Eggers_2008_IOP}. Without stretching, the breakup dynamics is determined by the balance between viscosity and surface tension~\cite{Pita_2012_PRL}, characterised by the Ohnesorge number, $Oh=\mu/\sqrt{\sigma R\rho}$, where $\mu$ is the dynamic viscosity, $\sigma$ the surface tension, and $\rho$ the fluid density. For $Oh<\mathcal{O}(0.01)$, filaments break into multiple droplets, whereas for $Oh\ge\mathcal{O}(1)$ viscous forces dominate and the filament retracts into a single droplet~\cite{Schulkes_1996}.

In many applications, including suspensions, inkjet printing, and additive manufacturing, liquid filaments contain suspended particles or impurities that introduce localized perturbations and modify the instability dynamics. Studies of particle-laden jets and suspensions containing many particles~\cite{CLARKE_2010,Zhang_2022_ACS,GRAMLICH_2012} have shown that particles alter breakup patterns and droplet-size distributions. Likewise, studies of dense suspensions~\cite{Morris_PRF_2004,Thievenaz_SM_2021,Thievenaz_PNAS_2022} demonstrated that multiple particles strongly influence the breakup of liquid drops and filaments. In dilute suspensions, however, localized perturbations generated by a single particle become important, raising the fundamental question of how an individual particle modifies the breakup dynamics and pinch-off time of a stretching filament.

In addition to particle-induced disturbances, stretching continuously decreases the filament radius while mass is lost through the ends at a rate $J_m=\rho u_{max}$, where $u_{max}$ is the characteristic velocity. The thinning rate is $\dot{\epsilon}=u_{max}/L$, defining the stretching time scale $\tau_s\sim1/\dot{\epsilon}$. Breakup is therefore governed by the competition between the viscous time scale $\tau_v$ and the stretching time scale $\tau_s$.

The importance of particle-induced perturbations can be characterised by $\beta(t)=r_p/R({\bf x}_p,t)$, where $r_p$ is the particle radius and $R({\bf x}_p,t)$ is the local filament radius. When $\beta(t)<1$, particle-induced perturbations are negligible. As stretching reduces $R$, however, $\beta$ increases, and once $\beta(t)\sim1$, a localized perturbation develops, depending on the particle size and the surrounding lubrication layer, which can subsequently trigger breakup. Despite the importance of suspended particles in filament breakup~\cite{Morris_PRF_2004,Thievenaz_SM_2021,Thievenaz_PNAS_2022}, the fundamental role of a single particle-induced localized perturbation in a stretching filament remains poorly understood.
 
\begin{figure}
    \centering
    \includegraphics[scale=0.3]{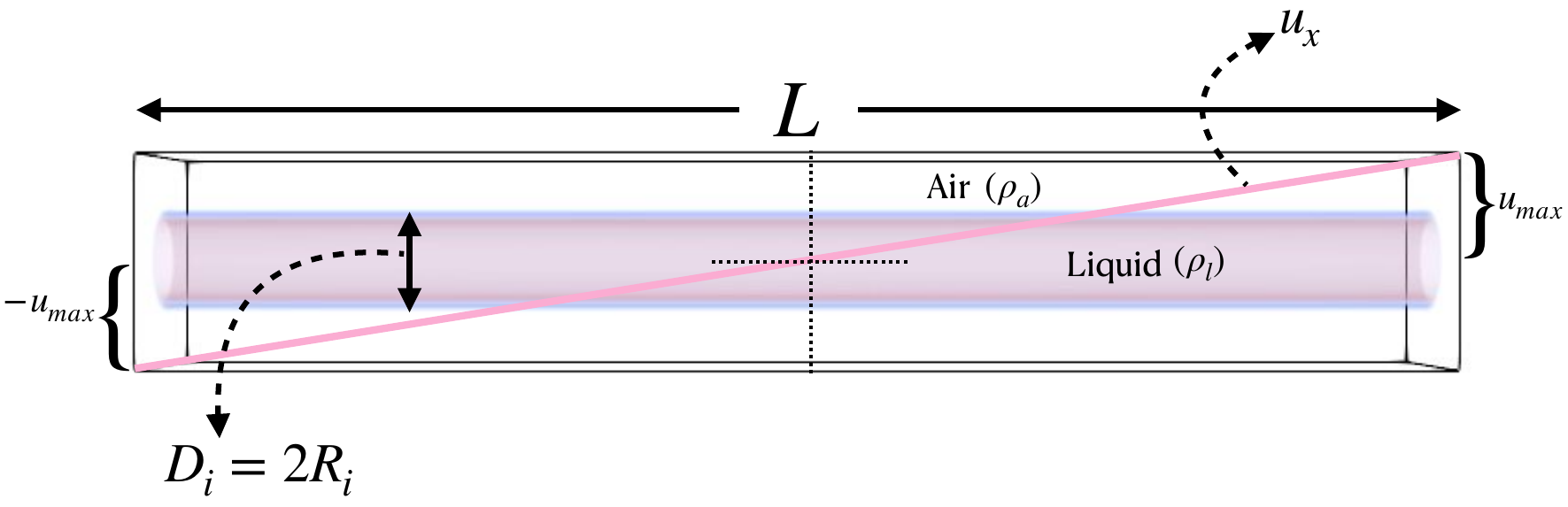}
    \caption{Setup of a liquid filament of density $\rho_l$, length $L$, and initial radius $R_i=R(t=0)$ with a solid particle (grey) at its center in a three-dimensional (3D) box. The solid pink line denotes the linear axial velocity profile, $u_x=2/L(x-L/2)u_{max}$, with $-u_{max}$ and $u_{max}$ imposed at the left and right ends, respectively. The surrounding fluid is air of density $\rho_a$.}
    \label{fig:filament_schematic}
\end{figure}

\begin{figure*}[!htb]
    \centering
    \includegraphics[width=1.0\linewidth]{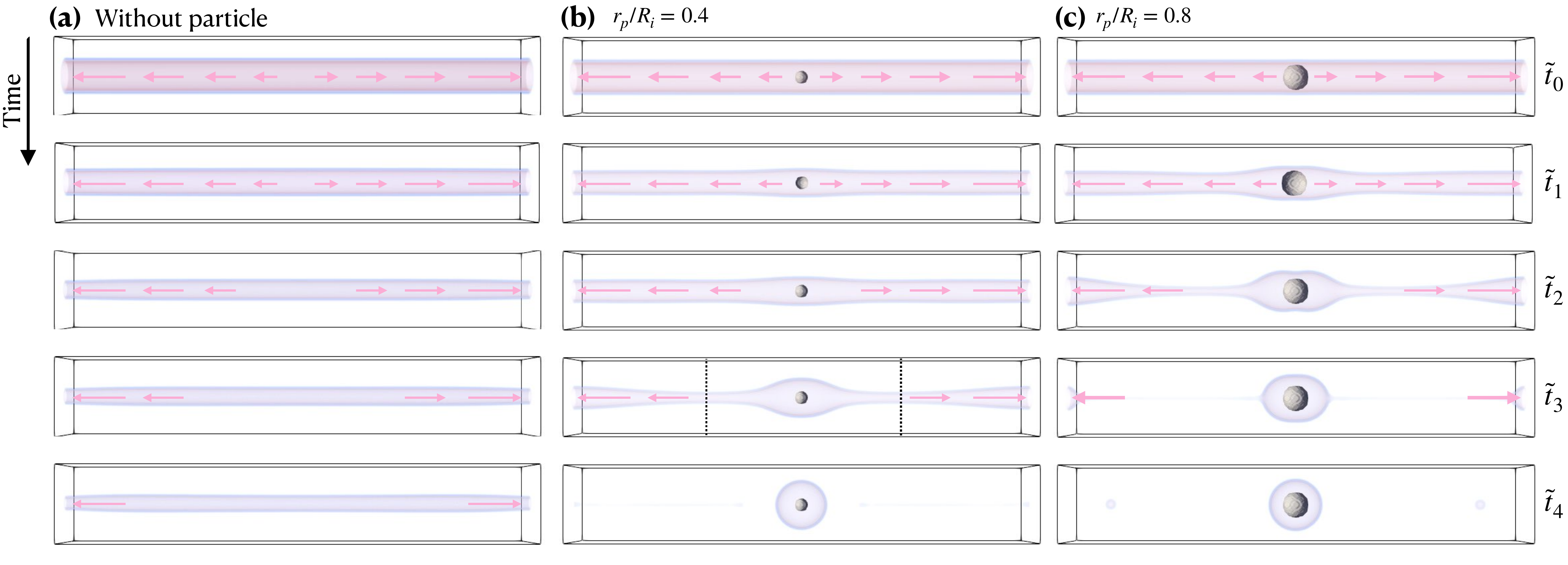} 
    \caption{Volume renderings of a stretching filament at five representative times without a particle in {\bf (a)} and with a solid particle of radius $r_p/R_i=0.4$ in {\bf (b)} and $0.8$ in {\bf (c)}, where $R_i$ is the initial filament radius. The imposed outflow rate is $\tilde{\dot{\epsilon}}=10.42\times10^{-3}$, corresponding to $Ca=\tilde{\dot{\epsilon}}L/R_i=0.3$. Pink arrows indicate the imposed linear axial velocity profile, with arrow length representing velocity magnitude. Time is non-dimensionalised as $\tilde{t}=t/\tau_v$, where $\tau_v=\mu R_i/\sigma$ is the capillary viscous time scale. The representative times are $\tilde{t}_0=0$, $\tilde{t}_1=25.67$, $\tilde{t}_2=36.67$, $\tilde{t}_3=42.53$, and $\tilde{t}_4=66.00$. The filament radius $R(x,t)$ decreases due to the imposed outflow, and the minimum radius $R_{\min}(t)$ is indicated by the dashed vertical lines.}
    \label{fig:lig_part_anim}
\end{figure*}

In this Letter, we show that a solid particle acts as a localized perturbation that triggers a universal self-similar pinch-off regime once $\beta(t)\sim1$. Using numerical simulations and analytical modelling, we demonstrate that the ensuing pinch-off is self-similar and independent of particle size. Existing experiments on particle-laden filaments have observed particle-induced accelerated pinch-off~\cite{Lindner_2015, Thievenaz_PNAS_2022}. However, the controlled minimal setup considered here provides a systematic framework to quantify particle-size dependence of the pinch-off time and to resolve the associated self-similar regime. We derive an analytical expression for the pinch-off time that captures the interplay between filament stretching and capillary instability and quantitatively agrees with the simulations.

To study the effect of particle-induced perturbations on filament breakup, we consider a cylindrical filament of density $\rho_l$, length $L$, and initial radius $R_i=R(t=0)$ (Fig.~\ref{fig:filament_schematic}). This provides a simple setup that closely mimics the hydrodynamic conditions of a generic stretching filament. A solid spherical particle of radius $r_p$ is placed at its centre with a no-slip boundary condition on its surface. Axial stretching is imposed by prescribing velocities $-u_{max}$ and $u_{max}$ at the left and right ends, respectively, producing the linear axial velocity profile $u_x=2/L(x-L/2)u_{max}$. Incompressibility requires transverse velocity components $u_{y,z}=-(u_{max}/L)(y,z)$, obtained from continuity. These are smaller than the axial velocity by the filament slenderness [$\sim\overline{R}(t)/L\approx0.035$ initially and decreasing with time]. In the simulations, the complete velocity field, including $u_{y,z}$, emerges self-consistently from the imposed $\pm u_{max}$ boundary conditions. Outside the filament, $\nabla\cdot{\bf u}=0$ is imposed at the left and right boundaries, with periodic boundary conditions elsewhere. This simple configuration isolates the effect of localized perturbations during filament thinning- an aspect that is difficult to probe in more complex geometries.

The computational domain is treated as an open system, allowing mass to leave at a prescribed outflow rate $\dot{\epsilon}=u_{max}/L$. As mass exits, the filament stretches and its radius decreases. For particles smaller than the initial filament radius, particle-induced perturbations are negligible during the early stages [$\beta(t)<1$], and the filament radius $R(x,t)$ evolves primarily through mass conservation. As stretching continues and $\beta(t)\sim1$, particle-induced perturbations become significant and ultimately govern the breakup dynamics. Since breakup is controlled by both stretching and capillary-viscous forces, we characterise the outflow rate by the dimensionless parameter $\tilde{\dot{\epsilon}}=\tau_v/\tau_s$, where $\tau_v=\mu R_i/\sigma$ is the capillary-viscous time scale.

Numerical simulations are performed using the lattice Boltzmann method (LBM) with the color-gradient (CG) model for two-phase immiscible flows in a domain of size $N_x\times N_y\times N_z$, where $N_x=432$ and $N_y=N_z=72$~\cite{Rosis_2020_POF,leclaire_2017_PRE}. The surrounding fluid is air with density $\rho_a$ and density ratio $\rho_l/\rho_a=1000$. The particle dynamics follows Newton's law with hydrodynamic forces obtained from two-way fluid--solid coupling. The filament aspect ratio is $L/R_i=28.8$. Governing equations and numerical details are provided in the Supplemental Material~\cite{supmat}. {Unless stated otherwise, all results presented below correspond to $Oh=1$; additional Ohnesorge numbers are reported in the Supplemental Material~\cite{supmat} to demonstrate the robustness of the self-similar regime.

Figure~\ref{fig:lig_part_anim} shows the evolution of a stretching filament without a particle in {\bf (a)} and with solid particles of radius $r_p/R_i=0.4$ and $0.8$ in {\bf (b)} and {\bf (c)}, respectively, where $R_i$ is the initial filament radius. For the representative case shown here, the dimensionless outflow rate is $\tilde{\dot{\epsilon}}=10.42\times10^{-3}$, corresponding to $Ca=\tilde{\dot{\epsilon}}L/R_i=0.3$ Later we explore a broad range of outflow rates and $Ca$]. The imposed outflow generates a linear axial velocity profile with zero velocity at the filament centre, as indicated by the purple arrows. To isolate particle-induced perturbations from particle motion, the particle is placed at the filament centre and remains stationary throughout the simulation. In the presence of a particle, the filament pinches off, with breakup occurring earlier for the larger particle [$r_p/R_i=0.8$, Fig.~\ref{fig:lig_part_anim}{\bf (c)}] than for the smaller one [$r_p/R_i=0.4$, Fig.~\ref{fig:lig_part_anim}{\bf (b)}]. Without a particle [Fig.~\ref{fig:lig_part_anim}{\bf (a)}], the filament thins monotonically over the same time interval without pinching off, highlighting the role of particle-induced perturbations in triggering necking. The absence of visible surface corrugations in Fig.~\ref{fig:lig_part_anim}{\bf (a)} results from the combined effects of moderate viscosity ($Oh=1$) and stretching. Because the base state evolves in time, stretching increases the perturbation wavelength and suppresses its visible growth over the timescale considered. Simulations at lower $Oh=0.67$ under the same stretching rate show faster perturbation growth, consistent with this interpretation (Supplemental Material~\cite{supmat}).

\par{\textit{Self-similar pinch-off time}-}
\begin{figure*}[!htb]
    \centering
    \includegraphics[width=1\linewidth]{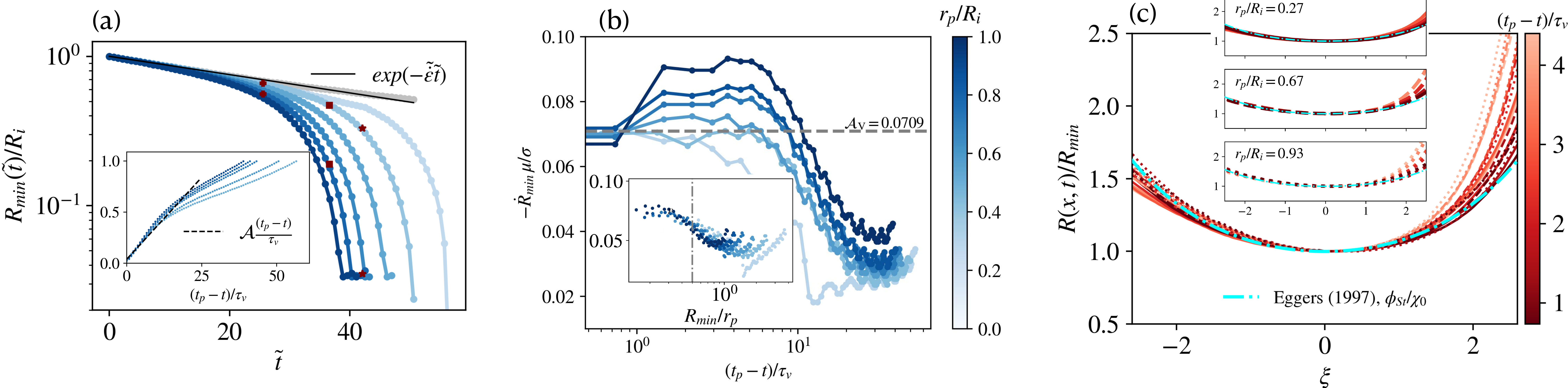}
    \caption{{\bf (a)} Normalised minimum filament radius $R_{\min}(t)/R_i$, where $R_{\min}(t)=\min\{R(x,t)\}$ and $R_i$ is the initial radius, versus scaled time $\tilde{t}=t/\tau_v$ for different particle sizes $r_p/R_i$. The inset shows $R_{\min}(t)/R_i$ versus $(t_p-t)/\tau_v$, demonstrating the collapse of all curves near pinch-off. Small upticks at the end of some curves are numerical artifacts arising when $R_{\min}$ approaches the simulation resolution and are not physical. {\bf (b)} Dimensionless local thinning rate $-\dot{R}_{min}\mu/\sigma$ versus scalted time $(t_p-t)$ for different particle sizes. The rate grows non-monotonically, reaches a maximum, and eventually converges to its asymptotic value of $\mathcal{A}_V=0.0709$ in the viscous regime. The inset shows the local thinning rate versus $R_{min}/r_p$ with a vertical dashed line at $R_{min}/r_p=0.5$. {\bf (c)} Interface profiles $R(x,t)/R_{min}$  versus the similarity variable $\xi$ at six times approaching pinch-off for $r_p/R_i=0.27$ (solid, $-$), $0.67$ (dahsed, $--$), and $0.93$ (dotted, $:$). Profiles collapse within the neck region. Close to pinch-off, the profile converges to Egger's self-similar solution $\phi_{St}$ [cyan curve]. The inset shows each particle size separately for clarity. Dark red markers in {\bf (a)} correspond to the snapshots in Fig.~\ref{fig:lig_part_anim}(b)--(c) at $\tilde{t}_1$ [$\textcolor{Maroon}{+}$], $\tilde{t}_2$ [$\textcolor{Maroon}{\blacksquare}$], and $\tilde{t}_3$ [$\textcolor{Maroon}{\star}$]. Subtracting $t_{exp}$ from $t_p$ yields the self-similar pinch-off time $t_{ps}$, which is independent of particle size.}
    \label{fig:rad_vs_time}
\end{figure*}
For particles smaller than the initial filament radius, the filament thins monotonically at early times [Fig.~\ref{fig:lig_part_anim}(b)]. As stretching proceeds and the filament surface approaches the particle [$\beta(t)\sim1$], the interface dynamics becomes controlled by the particle proximity, and the subsequent perturbation growth is independent of particle size. This marks the onset of a self-similar instability leading to pinch-off. We denote this time by $t_{exp}$ and quantify it below.

Before the filament reaches the particle [Fig.~\ref{fig:lig_part_anim}(b)–(c)], or in the absence of a particle [Fig.~\ref{fig:lig_part_anim}(a)], the filament radius follows directly from mass conservation, providing an estimate of $t_{exp}$. For a cylindrical filament of length $L$ and radius $R(t)$, the mass is $m(t)=\pi R(t)^2L\rho_l$. With outflow velocities $\pm u_{max}$ imposed at the ends, $\dot{m}(t)=-2\rho Au_{max}$, giving $\dot{R}/R=-u_{max}/L$. Solving this equation yields
\begin{eqnarray}
   \overline{R(t)} \equiv R(t) = R_i e^{-\tilde{\dot{\epsilon}} \tilde{t}}\,,
   \label{eq:radius_exponential}
\end{eqnarray}
\begin{figure}
    \centering
    \includegraphics[scale=0.17]{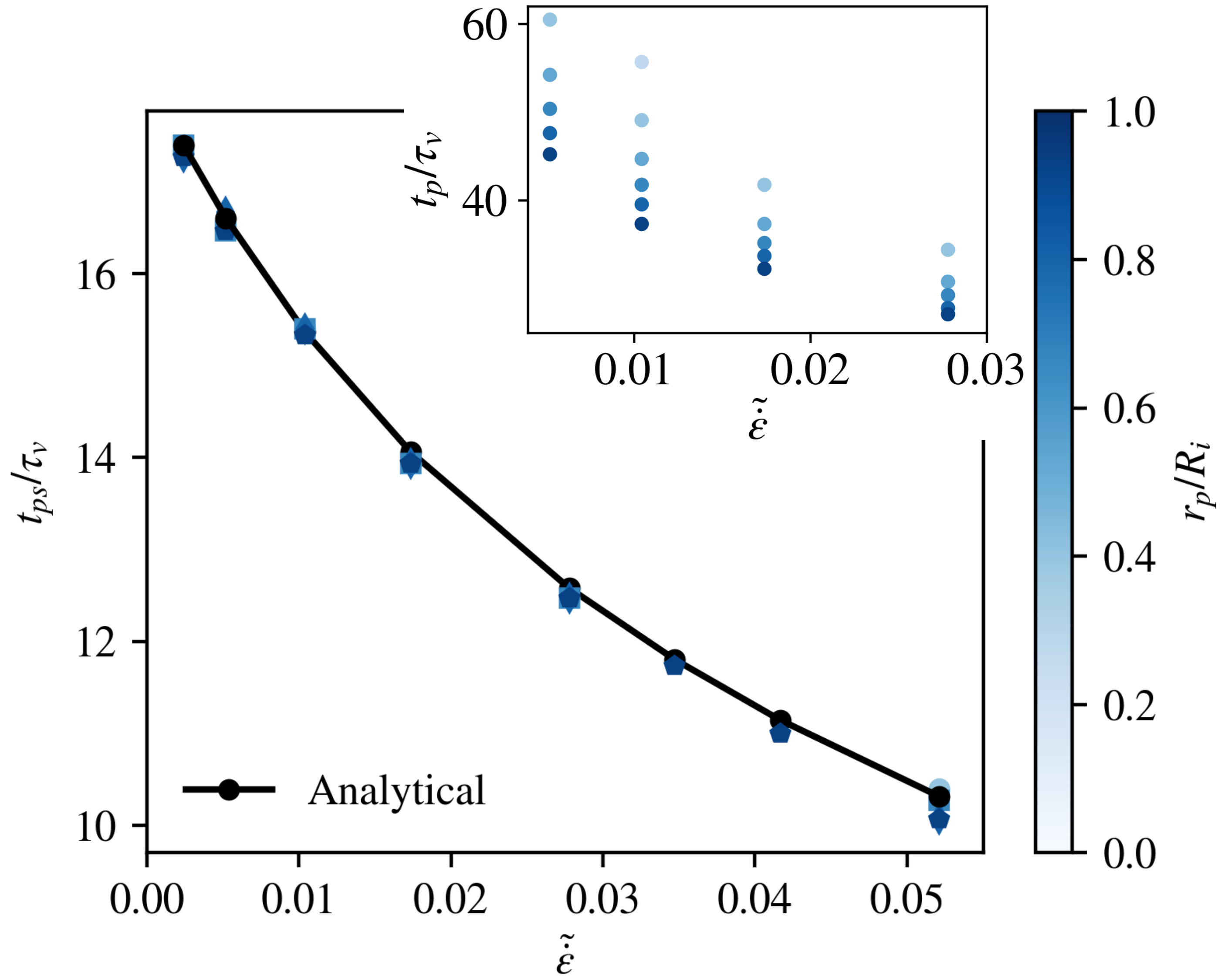}
    \caption{Self-similar pinch-off time $t_{ps}=t_p-t_{exp}$, normalised by $\tau_v$, versus outflow rate $\tilde{\dot{\epsilon}}$ for different particle sizes. The inset shows the total pinch-off time $t_p$ versus $\tilde{\dot{\epsilon}}$.}
    \label{fig:analytical_pinchoff}
\end{figure}
where $\tilde{\dot{\epsilon}}=(u_{max}/L)\tau_v$ is the dimensionless outflow rate, $\tilde{t}=t/\tau_v$, and $\tau_v=\mu R_i/\sigma$ is the capillary-viscous time scale. The exponential behaviour of Eq.~\eqref{eq:radius_exponential} also follows from the long-wavelength approximation for the unperturbed filament~\cite{Morris_2009}. We note that the pressure in this base state contains the curvature contribution $\gamma/\overline{R(t)}-2\mu\dot{\epsilon}$ (Supplemental Material~\cite{supmat}). The flow is driven by the imposed outflow rate $\dot{\epsilon}$, while the curvature contribution vanishes for the idealised base state. In the presence of a particle, however, necking generates curvature gradients that drive the localized perturbation.

To characterise the onset of self-similar pinch-off, we compute the minimum filament radius $R_{\min}(t)=\min{R(x,t)}$~\cite{Cohen_1999_PRL}, where $x$ is the axial coordinate [indicated by the dashed vertical lines in Fig.~\ref{fig:lig_part_anim}(b)]. Figure~\ref{fig:rad_vs_time}(a) shows $R_{\min}(t)$ for different particle sizes for the representative case $\tilde{\dot{\epsilon}}=10.42\times10^{-3}$ [See the Supplemental Material~\cite{supmat} for variation of $R_{\min}(t)$ for different resolutions]. Without a particle, $R_{\min}$ follows the exponential thinning predicted by Eq.~\eqref{eq:radius_exponential} (grey curve). With a particle, $R_{\min}$ initially follows the same exponential evolution $\overline{R(t)}$, corresponding to the unperturbed regime.

We emphasize that the evolution proceeds through distinct stages. Initially, the filament thins kinematically via mass conservation (Eq.~\eqref{eq:radius_exponential}), without a diverging quantity. At later times, the particle-induced perturbation grows until nonlinear effects drive the transition to the self-similar pinch-off regime, which follows $R_{min}/R_i = \mathcal{A}(t_p-t)/\tau_v$ (Fig.~\ref{fig:rad_vs_time}(a), inset), with time-dependent prefactor $\mathcal{A}$. Very close to the pinch-off, it is calculated to be $0.07$, observed in the classical viscous self-similar solution~\cite{Eggers_2008_IOP}. 

Since the approach of $R_{min}(t)$ to its scaling regime can be non-monotonic and prefactor $\mathcal{A}$ is time-dependent~\cite{Li_Sprittles_2016,lagarde_pnas_2018,Deblais_PRL_2018}, we calculate $\mathcal{A}$ using the dimensionless local thinning rate $-\dot{R}_{min}\mu/\sigma = \mathcal{A}$ and plot in Fig.~\ref{fig:rad_vs_time}(b) versus the scaled time $(t_p-t)/\tau_v$. For all the particle sizes considered, the rate $-\dot{R}_{min}\mu/\sigma$ shows the non-monotonic behaviour reaching a maximum value, which shows dependence on the particle size. The rate eventually converges, for all particle sizes, to its asymptotic value $\mathcal{A}_{\rm V} = 0.0709$ for the viscous regime~\cite{Corpart_Herrada_Deblais_Bonn_2025} within $\approx[5]\%$. Plotted against $R_{min}/r_p$ in the inset of Fig.~\ref{fig:rad_vs_time}(b), the rates for different particle sizes merge onto a single curve once $R_{min}/r_p \approx [0.5]$ (vertical dashed line), marking the instance when particles do not affect the thinning.

We further characterize self-similarity using the full interface profile, $R(x,t)/R_{min}(t) = \phi(\xi)$, using the similarity variable $\xi=l/t'^{\beta}$, where $l = (x-x(R_{min}))/l_\mu$, $t' = (t_p-t)/\tau_{\mu}$, and the viscous exponent $\beta = 0.175$~\cite{Eggers_RevModPhys_1997}. Here, $l_\mu=\mu^2/(\sigma\rho_l)$ and $\tau_{\mu} = \mu^3/(\rho_l \sigma^2)$ are the characteristic length and time scales. Figure~\ref{fig:rad_vs_time} (c) shows $R(x,t)/R_{min}$ at six times approaching $t_p$ for the three particle sizes. Near the neck, profiles from different times and different particle sizes collapse onto a single curve, demonstrating that the late-stage breakup dynamics are independent of particle sizes. We also overlay the self-similar solution $\phi(\xi)_{St}/\chi_0$ [dashed cyan curve in Fig.~\ref{fig:rad_vs_time} (c)] obtained by Eggers~\cite{Eggers_RevModPhys_1997} for a viscous thread. Earlier profiles are asymmetric, being steeper on the particle side, where the particle holds up the radius, while the other side thins under stretching. As pinch-off is approached, this asymmetry gradually disappears, and for $(t_p-t)/\tau_v \lesssim [2]$ the profiles for all particle sizes converge to the self-similar solution for $|\xi| \lesssim [3]$. The inset shows the profiles separated by particle size, highlighting the temporal collapse for each case.

The self-similar regime persists for $Oh\gtrsim1$: additional simulations at $Oh=0.67,1,1.2,1.4$ show the same profile collapse across particle sizes (Supplemental Material~\cite{supmat}). At the lower end of this range, however, the smallest particles no longer trigger the perturbation before spontaneous inertial-capillary breakup occurs, defining the limit of validity of the particle-induced mechanism. Self-similar profile collapse for different filament aspect ratios is also shown in the Supplemental Material~\cite{supmat}.

At later times, deviations from exponential thinning emerge due to the particle-induced perturbation. We define the onset time $t_{exp}$ as
\begin{eqnarray}
    t_{exp} = \min \left\{ t \;:\; \left| R_{min}(t) - \overline{R}(t) \right| \ge R_{\mathrm{th}} \right\}\,,
    \label{eq:t_exp}
\end{eqnarray}
where $R_{th}$ is a prescribed threshold of $5$--$10\%$ of $\overline{R}(t)$. This time-scale provides an estimate of when the filament surface first approaches the particle and the perturbation begins to grow. 

The pinch-off time $t_p$ is defined as the instant when $R_{min}(t)$ falls below the spatial resolution of the simulation. The self-similar pinch-off time is then $t_{ps}=t_p-t_{exp}$, representing the duration of particle-induced necking. Figure~\ref{fig:analytical_pinchoff} shows $\tilde{t}_{ps}$ versus the outflow rate $\tilde{\dot{\epsilon}}$ for different particle sizes. For a given $\tilde{\dot{\epsilon}}$, $\tilde{t}_{ps}$ collapses onto a single value for all particle sizes, demonstrating that the perturbation growth and resulting pinch-off dynamics are independent of particle size. In contrast, the total pinch-off time $t_p$ retains a particle-size dependence (inset of Fig.~\ref{fig:analytical_pinchoff}). This collapse provides evidence for a universal self-similar regime governing late-stage breakup.\\

\par{\textit{Analytical prediction of the self-similar pinch-off time}-} The self-similar regime begins once the filament surface approaches the particle. A localized perturbation then develops at the filament centre, while the remainder of the filament continues to thin according to the unperturbed evolution, $\overline{R}(t)$, given by Eq.~\eqref{eq:radius_exponential}. We therefore write $R(x,t) = \overline{R(t)} + \delta R(x,t)$, where $\delta R$ is the particle-induced perturbation. Assuming axial symmetry, we express the perturbation as $\delta R(x,t) = a(t) f(x)$, where $f(x)$ captures the spatial structure imposed by the particle, and $a(t)$ determines its temporal growth.

Once the perturbation is introduced, its growth follows the universal dynamics of the Rayleigh-Plateau instability, independent of particle size. At this stage and for a single particle, the functional form of $f(x)$ and its dependence on particle can be neglected. The perturbation amplitude therefore evolves as  $a(t) = a_0 \exp{(\int_0^t \lambda (t')dt')}$, where $\lambda$ is the growth rate and $a_0$ is the initial amplitude of the perturbation. The growth rate $\lambda(t) = \sigma/(\mu \overline{R}(t))$ is defined through the base radius $\overline{R}(t)$ rather than particle size $r_p$ itself, so the functional form of the growth law is independent of particle size by construction. Treating the growth as a simple exponential in time, $a(t) = a_0 e^{\lambda t}$, requires that the rate at which $\overline{R}(t)$ changes is slow compared to the perturbation growth rate, i.e., a quasi-static approximation [see the Supplemental Material~\cite{supmat}]: $\lambda'(t)/\lambda(t)^2= \dot\epsilon/\lambda \ll 1$. In dimensionless form this reads $\tilde{\dot\epsilon} \ll R_i/\overline{R}(t)$, and evaluated at $t_{exp}$, where $\overline{R}(t_{exp}) \sim r_p$, becomes $\tilde{\dot\epsilon} \ll R_i/r_p$. Across the particle sizes and outflow rates considered, $R_i/r_p$ ranges from $1.08$ to $3.7$ while $\tilde{\dot\epsilon}$ ranges from $0.0024$ to $0.04$; the quasi-static condition therefore holds throughout. The interface profile can then be written as
\begin{eqnarray}
    R(x,t) = \overline{R(t)} + a(t) f(x)\,.
\end{eqnarray}

The pinch-off occurs when the perturbation amplitude becomes comparable to the mean radius, $a(t) \sim \overline{R(t)}$, which after simplifying yields the characteristic pinch-off time $ t_{p,a} = 1/(\lambda + \dot{\epsilon}) \ln \left(R_i/a_0\right)$. For the Ohnesorge number $Oh \sim 1.0$ and close to the surface of the particle, viscous effects dominate the dynamics. In this regime, we identify $\lambda$ as the capillary-viscous time scale given as $\lambda = \sigma/(\mu \overline{R(t)})$, where $\sigma$ and $\mu$ denote the surface tension and viscosity, respectively. Using Eq.~\eqref{eq:radius_exponential}, we get the pinch-off time scale as follows
\begin{eqnarray}
    t_{p,a} = \frac{1}{\lambda_0 e^{\dot{\epsilon}t_{p,a}} + \dot{\epsilon}} \ln \left(\frac{R_i}{a_0}\right)\,,
    \label{eq:analytical_pinchoff}
\end{eqnarray}
where $\lambda_0 = \sigma/(\mu R_i)$. In the absence of stretching ($\dot{\epsilon} = 0$), this expression reduces to the viscous capillary time scale $t_{p,a} \sim 1/\lambda_0$, corresponding to the breakup of a stationary filament with a small perturbation. It is important to note that near the particle surface, the dominant effect comes from viscosity. Therefore, the time scale $t_{p,a}$ provides a universal scaling for Ohnesorge numbers  $Oh\sim 1$, irrespective of other parameters such as the range of outflow rates considered in our simulations and the aspect ratio $L/R$ of the filament. We note that ``near" refers to the physical gap $\delta(t) = R(\mathbf{x}_p,t) - r_p$ becoming comparable to the threshold $R_{th}$ used to define $t_{exp}$ in Eq.~\eqref{eq:t_exp}, i.e.\ once the filament radius at the particle location deviates from the unperturbed exponential thinning of Eq.~\eqref{eq:radius_exponential}.

In the analytical expression for $t_{p,a}$ given by Eq.~\eqref{eq:analytical_pinchoff}, the only free parameter is the relative amplitude of the perturbation $a_0/R_i$. We determine this parameter by considering the case of the largest particle ($r_p/R_i=0.93$), where the particle is initially close to the filament surface and introduces a small perturbation consistent with the assumption of the analytical model. 

We then solve Eq.~\eqref{eq:analytical_pinchoff} iteratively for different outflow rates $\tilde{\dot{\epsilon}}$ to obtain the analytical pinch-off time $t_{p,a}$. Figure~\ref{fig:analytical_pinchoff} compares the time $t_{p,a}$ (solid black curve) with the self-similar pinch-off time $t_{ps}$ obtained from simulations. The agreement is excellent across all particle sizes, confirming that once the filament surface approaches the particle, the subsequent growth of perturbations follows a universal self-similar dynamics. This agreement holds over more than a decade in $\tilde{\dot{\epsilon}}$, across which $t_{ps}$ shows pronouned curvature. Expressed in terms of the Capillary number $Ca = \tilde{\dot{\epsilon}}L/R$, where $Ca= 0.07-1.53$ in our simulations, the results demonstrate that, for $Oh \sim 1$, the analytical time scale $t_{p,a}$ captures a universal scaling regime over the range of $Ca$ considered here.

In conclusion, we have developed a framework to investigate particle-induced localized perturbations during the breakup of a thinning liquid filament, providing a prototype for complex interfacial systems where surface tension, contact-line dynamics, and viscosity interact with solid boundaries. For particle sizes smaller than the initial filament radius, the filament initially evolves by mass conservation. Once the filament surface approaches the particle, the ensuing perturbation evolves toward a self-similar pinch-off regime that is independent of particle size. An analytical model predicts the self-similar pinch-off time in quantitative agreement with the simulations. Expressed in terms of the Capillary number, the model captures the universal scaling of the self-similar pinch-off time for $Oh\gtrsim1$ over the range of parameters considered.

In our setup, the final pinch-off of a stretching filament is governed by the competition between two characteristic time scales. The first is the stretching time scale, $t_s$, which is determined by the imposed outflow rate $\tilde{\dot{\epsilon}}$. The second is the time scale associated with the growth of the Rayleigh–Plateau (RP) instability. For $Oh \ll 1$, the growth of the RP instability is dominated by inertia and is characterized by the capillary time scale $t_{\sigma}  = \sqrt{(\rho R^3)/\sigma}$, whereas for $Oh\sim1$, viscous effects dominate and the relevant time scale becomes the capillary–viscous time $t_{\mu} = (\mu R)/\sigma$. When the stretching rate is not large enough, filaments with $Oh\ll1$ undergo breakup independently of the presence of particles because the instability grows rapidly on the short capillary time scale $t_{\sigma}$. In contrast, for $Oh\sim1$, the slower capillary–viscous time scale $t_{\mu}$ allows particle-induced perturbations to persist and grow, making their influence on the pinch-off dynamics significant even for relatively small particles. As a result, the late-stage breakup evolves toward a universal self-similar regime, leading to a pinch-off time that becomes independent of particle size.

In the present study, we have focused on the symmetric configuration in which the particle is placed at the center of the filament and remains stationary. This provides a minimal setting to isolate the effect of particle-induced perturbations. In more general situations, where the particle is off-center, it is advected by the flow and interacts dynamically with the thinning filament, leading to more complex breakup behaviour. The analytical framework developed here can be extended to account for such cases, which we leave for future work. 

\par{\textit{Acknowledgment}-} S.S. and F.T. thank Youssef Saade and Giovanni Soligo for useful discussions. This work is part of the FIP-2 research programme, (partly) financed by the Dutch Research Council (NWO). We further acknowledge support from the NWO project “Turbulent multiphase and rarefied gas flows” (file number 2025.008) for access to computational resources.


\section*{Supplemental Material}

\section{Governing Equations}
\label{sec:governing_equation}
 The dynamics of a stretching viscous ligament is given by the Navier-Stokes equations for an incompressible Newtonian Fluid given as follows
\begin{eqnarray}
    \frac{\partial {\bf u}}{\partial t} + {\bf u}\cdot \nabla {\bf u} &=& -\frac{1}{\rho_l}\nabla p + \nu \nabla^2 {\bf u}\,,\\
    \nabla\cdot {\bf u} &=&0\,,
    \label{eq:NS}
\end{eqnarray}
where ${\bf u}({\bf x},t)$ is the velocity field, $\rho_l$ is the density, and $\nu$ is the kinematic viscosity of the fluid. The translation and rotational dynamics of the inertial particle are governed by the Newtpon's law with the following equations
\begin{eqnarray}
    m_p\frac{d {\bf u}_p}{d t} &=& \oint_S {\bf \tau}_s \cdot {\bf n} dS\,,\\
    I_p\frac{d {\bf \omega}_p}{d t} &=& \oint_S {\bf r}\times ({\bf \tau}_s \cdot {\bf n}) dS\,,
    \label{eq:particle}
\end{eqnarray}
where $m_p$ and $I_p$ are the mass and moment of inertia of the particle, ${\bf \tau}_s = -p{\bf I} + \mu \nabla \cdot {\bf u}$ is the hydrodymanci stress tensor, ${\bf n}$ is the vector normal to the particle surface area $dS$, ${\bf u}_p$ and ${\bf \omega}_p$ are the velcolity and angular velocity of the particle.

The dynamics of the liquid ligament with the inertial particle follows the coupled equations given by Eq.~\eqref{eq:NS}-\eqref{eq:particle}, which do not posses the analytical solution. We solve them using numerical methods using the Lattice Boltzmann Method (LBM) based on the color gradient (CG) method~\cite{Rosis_2020_POF,leclaire_2017_PRE} for multicomponent fluids. 

\section{Color-gradient Lattice-Boltzmann Method}
\label{sec:cg_method}
The color-gradient (CG) is used to model multicomponent fluids using Lattice-Boltzmann Method (LBM). In our case one fluid is that of ligament and the other one is the air surrounding it. In LBM, each component is described by the discrete distribution function $f_i^k({\bf x}, t)$ at time $t$ and position ${\bf x}$, which evolves as follows
\begin{eqnarray}
    f_i^k({\bf x}+{\bf c}_i \Delta t, t+\Delta t) = f_i^k({\bf x}, t) + \Omega_i^k(f_i^k({\bf x}, t))\Delta t\,,
\end{eqnarray}
where $k$ denotes the fluid component (liquid or air). Here, ${\bf c}_i$ are the discrete velocities with $i=0,...,q-1$, describing the stencil or lattice, and $\Omega_i^k$ is the collision operator for each fluid. We use the D3Q19 lattice.

Macroscopic quantities of the fluids are obtained using the moments of the distribution function. The total density of each fluid is
\begin{eqnarray}
    \rho_k = \sum_i f_i^k ({\bf x}, t)\,,
\end{eqnarray}
with the total density given by $\rho({\bf x}, t) = \sum_k \rho_k $. The total momentum is given by 
\begin{eqnarray}
    \rho({\bf x}, t){\bf u}({\bf x}, t) = \sum_i f_i({\bf x}, t){\bf c}_i\,,
\end{eqnarray}
where $f_i({\bf x}, t) = \sum_k f_i^k({\bf x}, t)$.

In CG, the collision operator is a combination of three sub-operators given as follows
\begin{eqnarray}
    \Omega_i^k = \Omega_{3,i}^k(\Omega_{1,i}^k+\Omega_{2,i}^k)\,,
\end{eqnarray}
where the single-phase collisions are given by the BGK collision operator given by
\begin{eqnarray}
    \Omega_{1,i}f_i({\bf x}, t) = -\frac{1}{\tau}(f_i({\bf x}, t) -f_i^{eq}(\rho,{\bf u}))\,,
\end{eqnarray}
where $\tau$ is the relaxation time that determines the kinematic viscosity given as follows
\begin{eqnarray}
    \nu = \left(\tau-\frac{1}{2}\right)c_s^2\,,
\end{eqnarray}
with $c_s^2$ as the sound speed. Here, $f_i^{eq}$ is the equilibrium distribution function given as follows
\begin{eqnarray}
    f_i^{eq}({\bf x}, t) = \rho w_i \left(1+ \frac{3}{c^2}({\bf c}_i\cdot {\bf u})+\frac{9}{2c^4}({\bf c}_i\cdot {\bf u})^2-\frac{3}{2c^2}({\bf u}\cdot {\bf u})\right)\,,
\end{eqnarray}
where $w_i$ are the lattice weights~\cite{Rosis_2020_POF}.

The operator $\Omega_{2,i}^k$ is called the perturbation operator, which adds a color-gradient force to include the surface tension at the interface. The recoloring operator $\Omega_{3,i}^k$ redistributes the populations to maximise the immiscibility. See Ref.~\cite{leclaire_2017_PRE} for details on different operators and their implementation.

After the collision, the time evolution is performed using streaming, given as follows
\begin{eqnarray}
    f_i^k({\bf x}+{\bf c}_i \Delta t, t+\Delta t) = f_i^k({\bf x}, t)
\end{eqnarray}

The time evolution of translational dynamics of the particle is given by the following equation

\begin{eqnarray}
     m_p\frac{d {\bf u}_p}{d t} &=& m_l\frac{d {\bf u}}{d t} - \sum_k F_k \Delta V_p\,,
\end{eqnarray}
where $m_l = \rho_l V_p$ is the mass of the fluid displaced by the particle of volume $V_p$, and the second term accounts for the body forces on the particle.

For the rotational dynamics, we use a similar equation given as follows
\begin{eqnarray}
    I_p\frac{d {\bf \omega}_p}{d t} &=& I_f\frac{d {\bf \omega}}{d t}-\sum_k ({\bf x}_k-{\bf x}_c)\times F_k \Delta V_p\,,
\end{eqnarray}
where $ I_f$ is the moment of inertia of the fluid displaced by the particle, and ${\bf x}_c$ is its center of mass with ${\bf x}_k$ points on the surface of the particle.

\section{Plots of minimum radius for different resolutions}
\label{sec:minimum_radius}
Figure~\ref{fig:res_check} shows the plots of the normalised minimum filament radius $R_{\min}(t)/R_i$, where $R_{\min}(t)=\min\{R(x,t)\}$ and $R_i$ is the initial radius, versus scaled time $\tilde{t}=t/\tau_v$ for different resolutions,  $\Delta x$ (in red) and $0.5\Delta x$ (in green), where $\Delta x/R_i = 0.06$, showing the convergence of pinch-off time $t_p$.
\begin{figure}
    \centering
    \includegraphics[scale=1]{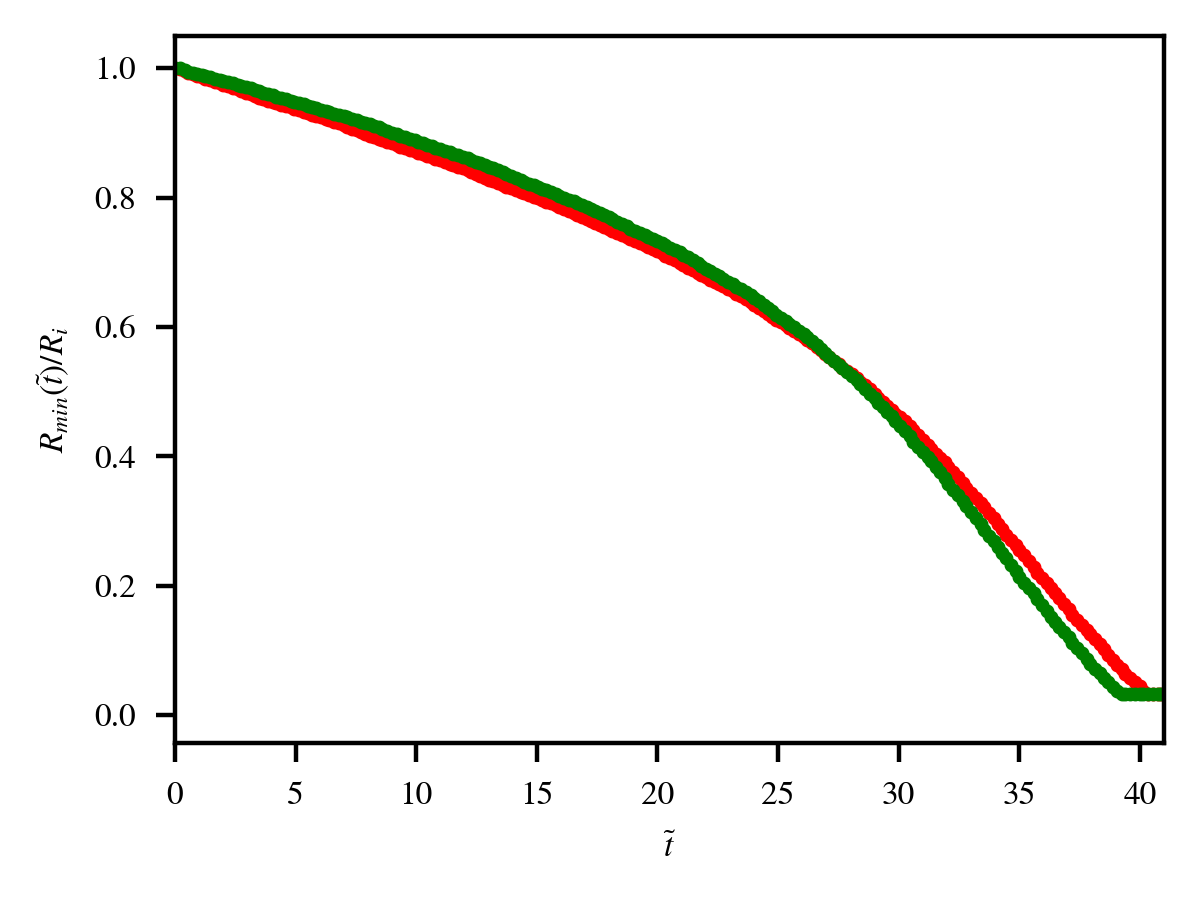}
    \caption{Normalised minimum filament radius $R_{\min}(t)/R_i$, where $R_{\min}(t)=\min\{R(x,t)\}$ and $R_i$ is the initial radius, versus scaled time $\tilde{t}=t/\tau_v$ for different resolutions,  $\Delta x$ (in red) and $0.5\Delta x$ (in green), where $\Delta x/R_i = 0.06$. }
    \label{fig:res_check}
\end{figure}

\section{Simulation for different $Oh$ number}
\label{sec:different_OH}
In this section, we present the results from our DNSs for different values of the Ohnesorge numbers $Oh$.
\subsection{Without the particle for $Oh=0.67$}
Figure~\ref{fig:anim_noparticle_o0.67} shows the volume renderings of a stretching ligament at five representative times without a particle for $Oh=0.6$ in comparison to the case in Fig.2(a) in the main text. This shows the faster growth of Rayleigh–Plateau instability at lower viscosity.
\begin{figure}
    \centering
    \includegraphics[scale=0.3]{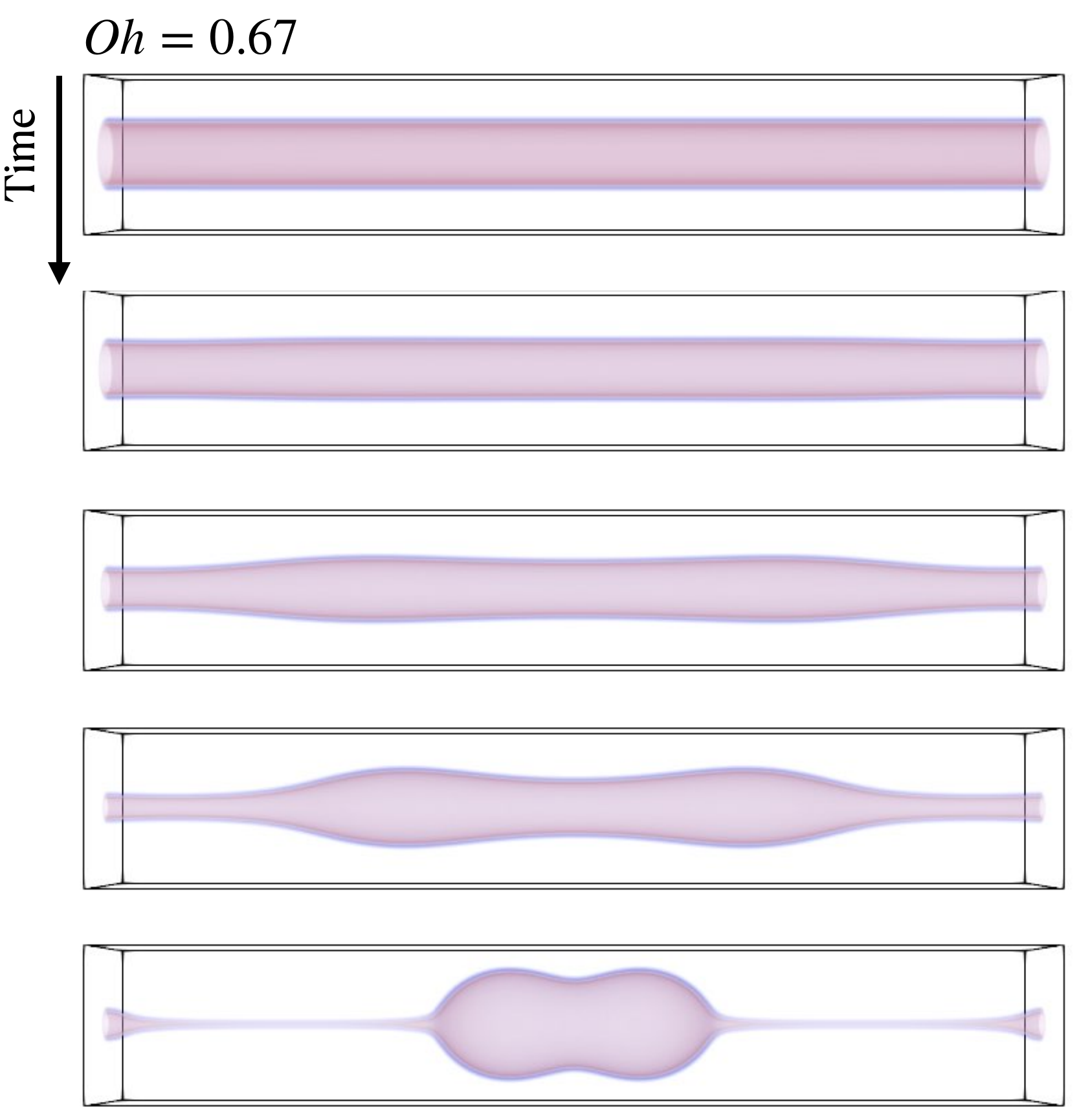}
    \caption{Volume renderings of a stretching ligament at five representative times, $\tilde{t} = 33.00, 49.50,57.75,66.00$, without a particle at $Oh=0.67$.}
    \label{fig:anim_noparticle_o0.67}
\end{figure}

\subsection{With the particle for $Oh=0.67$}
Figure~\ref{fig:anim_particle_o0.67} shows the volume renderings for $Oh=0.67$ and for particle sizes $r_p/R_i=0.26$ in (a) and $r_p/R_i=0.67$ in (b). Figure~\ref{fig:anim_particle_o0.67}(c) shows the self-similar profiles $\phi(\xi)$ versus the similarity variable $\xi$ at five times approaching pinch-off time, $t_p$, for the three particle sizes considered. Near the neck, profiles from different times and different particle sizes collapse onto a single curve, demonstrating that the late-stage breakup dynamics are self-similar and independent of particle size.

\begin{figure*}[!hbt]
    \centering
    \includegraphics[scale=0.13]{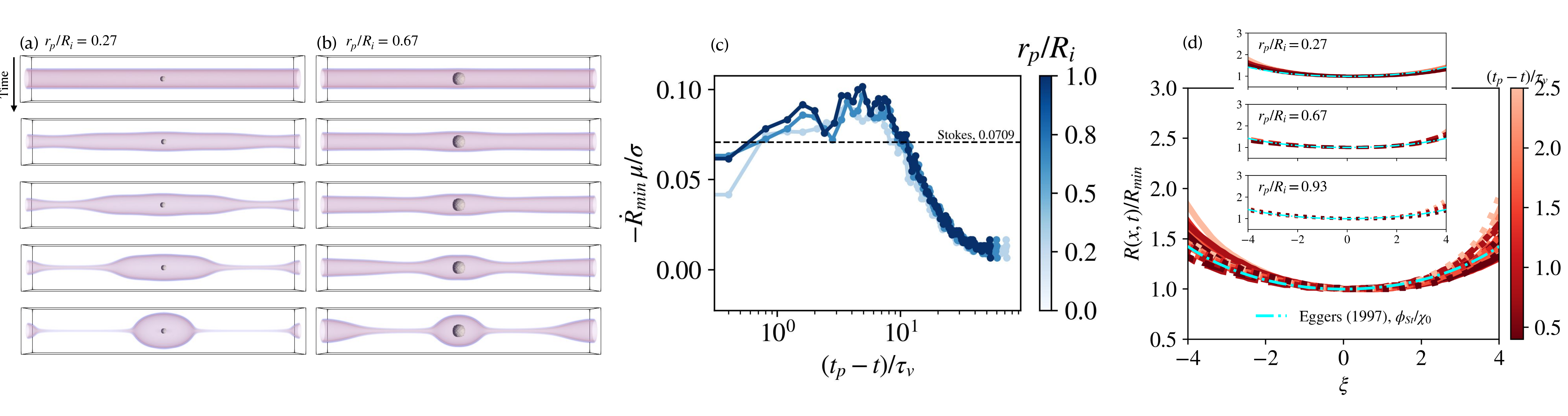}
    \caption{Volume renderings of a stretching ligament at five representative times with a solid particle of radius  $r_p/R_i = 0.27$ in {\bf (a)} and $r_p/R_i = 0.67$ in {\bf (b)}. {\bf (c)} Self-similar interface profiles, $\phi(\xi)$, plotted as a function of the similarity variable $\xi$ at five representative times $t'$ approaching pinch-off for particle sizes $r_p/R_i=0.27$ (purple), $r_p/R_i=0.67$ (red), and $r_p/R_i=0.93$ (green). Profiles for different particle sizes and times collapse within the neck region. The inset displays the profiles for each particle size in separate panels for clarity.}
    \label{fig:anim_particle_o0.67}
\end{figure*}

\subsection{With the particle for $Oh=1.2$}
Figure~\ref{fig:anim_particle_o1.2} shows the volume renderings for $Oh=1.2$ and for particle size $r_p/R_i=0.67$ in (b). Figure~\ref{fig:anim_particle_o0.67}(b) shows the self-similar profiles $\phi(\xi)$ versus the similarity variable $\xi$ at five times approaching pinch-off time, $t_p$, for the three particle sizes considered. Near the neck, profiles from different times and different particle sizes collapse onto a single curve, demonstrating that the late-stage breakup dynamics are self-similar and independent of particle size.
\begin{figure*}[!hbt]
    \centering
    \includegraphics[scale=0.17]{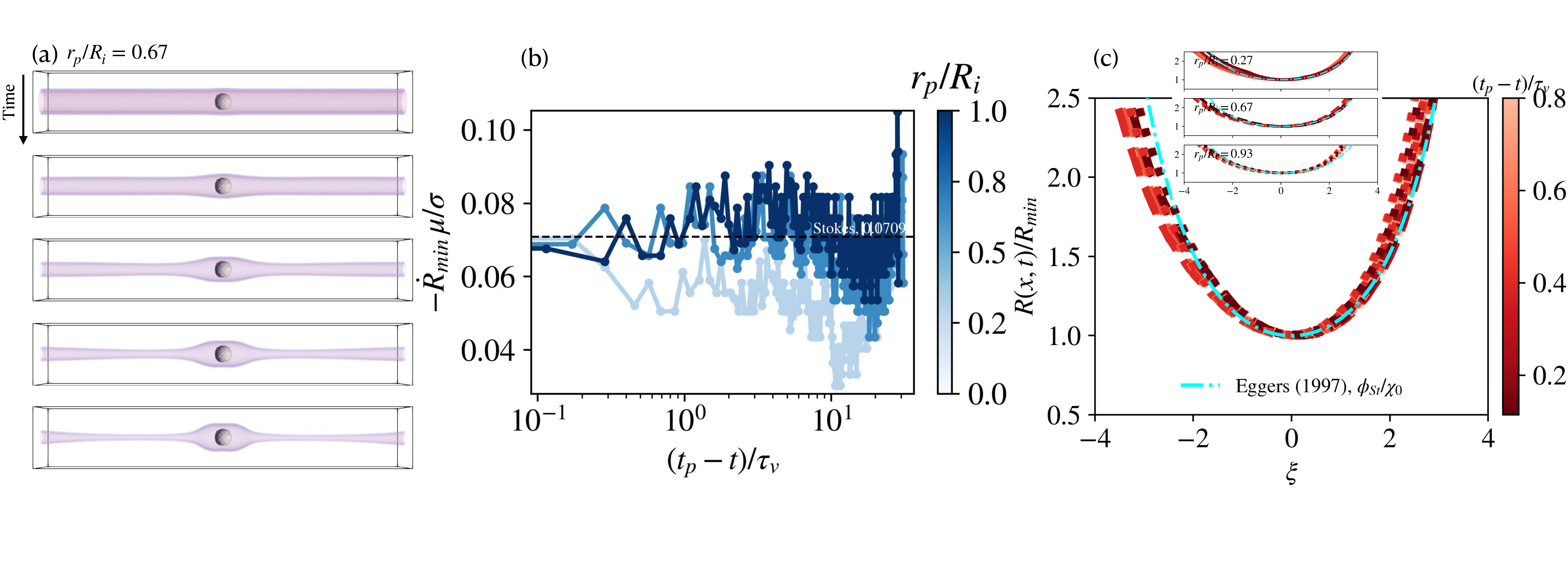}
    \caption{Volume renderings of a stretching ligament at five representative times with a solid particle of radius  $r_p/R_i = 0.67$ in {\bf (b)}. {\bf (b)} Self-similar interface profiles, $\phi(\xi)$, plotted as a function of the similarity variable $\xi$ at five representative times $t'$ approaching pinch-off for particle sizes $r_p/R_i=0.27$ (purple), $r_p/R_i=0.67$ (red), and $r_p/R_i=0.93$ (green). Profiles for different particle sizes and times collapse within the neck region. The inset displays the profiles for each particle size in separate panels for clarity.}
    \label{fig:anim_particle_o1.2}
\end{figure*}

\subsection{With the particle for $Oh=1.4$}
Figure~\ref{fig:anim_particle_o1.4} shows the volume renderings for $Oh=1.4$ and for particle size $r_p/R_i=0.67$ in (b). Figure~\ref{fig:anim_particle_o0.67}(b) shows the self-similar profiles $\phi(\xi)$ versus the similarity variable $\xi$ at five times approaching pinch-off time, $t_p$, for the three particle sizes considered. Near the neck, profiles from different times and different particle sizes collapse onto a single curve, demonstrating that the late-stage breakup dynamics are self-similar and independent of particle size.
\begin{figure*}[!hbt]
    \centering
    \includegraphics[scale=0.2]{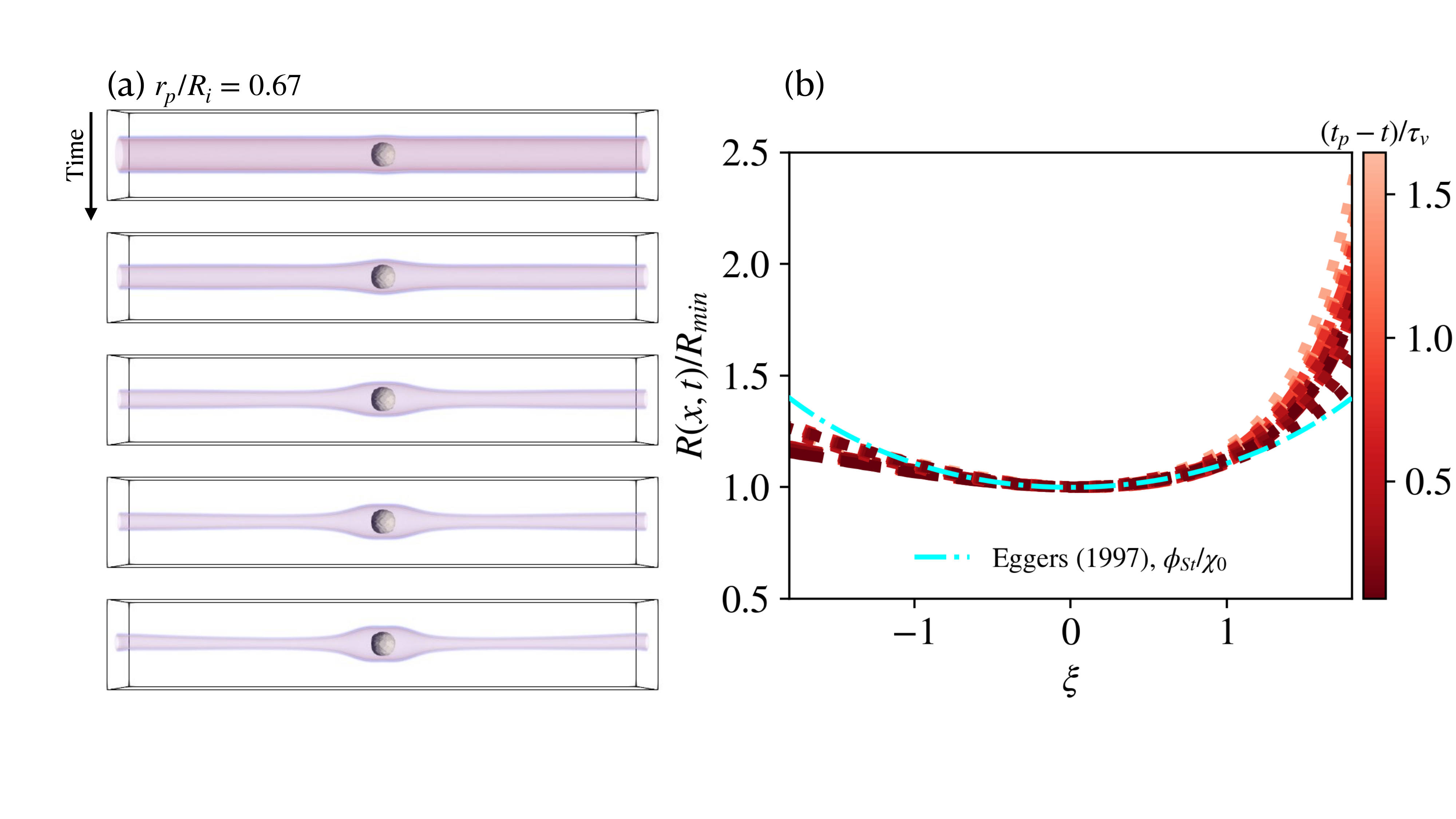}
    \caption{Volume renderings of a stretching ligament at five representative times with a solid particle of radius  $r_p/R_i = 0.67$ in {\bf (b)}. {\bf (b)} Self-similar interface profiles, $\phi(\xi)$, plotted as a function of the similarity variable $\xi$ at five representative times $t'$ approaching pinch-off for particle sizes  $r_p/R_i=0.67$ (purple), and $r_p/R_i=0.93$ (red). Profiles for different particle sizes and times collapse within the neck region. The inset displays the profiles for each particle size in separate panels for clarity.}
    \label{fig:anim_particle_o1.4}
\end{figure*}

\subsection{Self-similar profiles for different aspect ratios}
\begin{figure}
    \centering
    \includegraphics[scale=1]{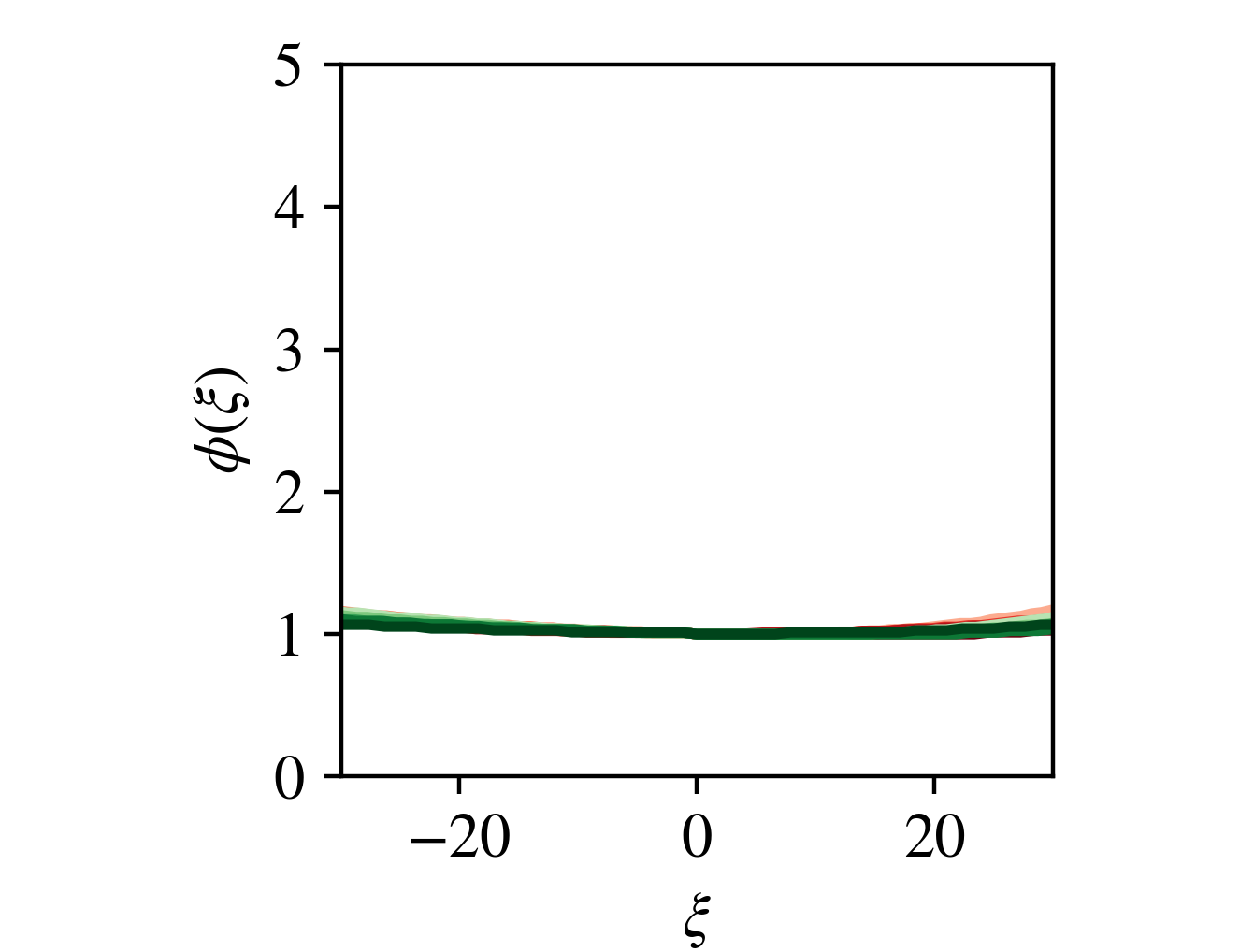}
    \caption{Self-similar interface profiles, $\phi(\xi)$, plotted as a function of the similarity variable $\xi$ at five representative times $t'$ approaching pinch-off for aspect ratio $L/R_i=57.6$ (red), and $L/R_i=86.4$ (green). Profiles for different aspect ratios and times collapse within the neck region. The particle size is $r_p/R_i=0.67$.}
    \label{fig:aspect_ratio}
\end{figure}

\section{Exponential thinning: Momentum conservation}
\label{sec:exponential_thinning}
We write the Navier-Stokes equation as follows:
\begin{eqnarray}
    \rho (\partial_t {\bf u} + {\bf u} \cdot \nabla {\bf u}) = -\nabla p + \mu \nabla^2 {\bf u}\,.
\end{eqnarray}

\begin{equation}
\begin{aligned}
\rho\left(\partial_t u_r
+u_r \partial_r u_r
+u_z \partial_z u_r\right)
&= -\partial_r p \\
&\quad +\eta\left(
\partial_r^2 u_r+\partial_z^2 u_r
+\frac{\partial_r u_r}{r}
-\frac{u_r}{r^2}\right),
\\[6pt]
\rho\left(\partial_t u_z
+u_r \partial_r u_z
+u_z \partial_z u_z\right)
&= -\partial_z p \\
&\quad +\eta\left(
\partial_r^2 u_z+\partial_z^2 u_z
+\frac{\partial_r u_z}{r}\right).
\end{aligned}
\end{equation}
Using the radial equation for $u_x = 2\dot{\epsilon}(x-L/2)$, and the corresponding radial velocity field $u_r = -\dot{\epsilon} r$:
\begin{eqnarray}
    \rho(\dot{\epsilon}^2 r) &=& -\partial_r p \,,
\end{eqnarray}
which gives $p = -\rho \dot{\epsilon}^2 r^2/2 + f(x,t)$. The free surface condition at $r=\overline{R}$ gives the pressure $p = -\rho \dot{\epsilon}^2 (r^2-\overline{R}^2)/2 + \gamma/\overline{R} - 2\mu \dot{\epsilon} + g(t)$. The axial momentum equation is also satisfied in the viscous regime.

\section{Base flow and perturbation: quasistatic approximation}
\label{sec:base_flow_plus_perturbation}
For a long filament, the linear axial velocity profile resulting from the stretching is $u_x = 2\dot{\epsilon}(x-L/2)$. If we use $\zeta =x-L/2$, we get $u(\zeta) = 2\dot{\epsilon}\zeta$. Using this, we get the material trajectories $d\zeta/dt = 2\dot{\epsilon}\zeta \to \zeta(t) = \zeta(0) \exp(2\dot{\epsilon} t)$. We can then define the Lagrangian label as $\eta = \zeta \exp(-2\dot{\epsilon} t)$. Using this label, we can write the material derivative for a field $f(\zeta, t)$
\begin{eqnarray}
    \left(\frac{\partial}{\partial t} + u\frac{\partial}{\partial \zeta}\right)f = \frac{\partial}{\partial t} f_{\eta}\,.
\end{eqnarray}
This suggests that we can choose the perturbation of the form
\begin{eqnarray}
    \delta R(\zeta,t) = \hat{A}(t) e^{ik_o \eta}; \delta u(\zeta,t) = \hat{B}(t) e^{ik_o \eta}\,.
\end{eqnarray}
We write the mass equation in the conservative form as follows
\begin{eqnarray}
    \partial_t (R^2) + \partial_x (R^2 u) = 0\,.
\end{eqnarray}
Linearizing the above equation using $R= \overline{R}+\delta R$, and $u = u_x+\delta u$, we get
\begin{eqnarray}
    \partial_t\hat{A} = -\dot{\epsilon} \hat{A}- \frac{ik(t) \overline{R}}{2}\hat{B}\,,
\end{eqnarray}
 $k(t) = k_0 e^{-2\dot\epsilon t}$ is the instantaneous wavenumber, diluted by the base-flow stretching. This identifies two distinct contributions: a geometric dilution at rate $\dot\epsilon$, common to any perturbation riding on the stretching base flow, and a capillary-driven growth term requiring the momentum equation for closure.\\
Closing this relation to obtain $\hat B$ in terms of $\hat A$ -- and thereby a fully derived growth rate $\lambda(t)$ in $a(t)=a_0e^{\int_0^t \lambda(t') dt'}$ -- requires solving the linearized momentum equation for a stretching base flow. The rigorous closure requires solving the full linearized axisymmetric Navier-Stokes equations with radial structure [due to a particle of finite size]. In lieu of the full closure, we justify the growth scaling law used in Eq.~(4) [in the main text] via a quasi-static approximation, which we verify holds throughout our parameter range $\tilde{\dot\epsilon} \ll R_i/r_p$, evaluated at $t_{exp}$.

\end{document}